\documentclass[pra,aps,twocolumn,10pt]{revtex4-1}
\usepackage{amssymb,amsmath,graphicx}

\begin{document}

\title{Gaussian beams diffracting in time}

\author{Miguel A. Porras}

\affiliation{Grupo de Sistemas Complejos, ETSIME, Universidad Polit\'ecnica de Madrid, Rios Rosas 21, 28003 Madrid, Spain}

\begin{abstract}
We show how to transform the mathematical expression of any monochromatic paraxial light beam into the expression of a pulsed beam whose diffraction is switched from the axial direction to its temporal structure. We exemplify this transformation with time-diffracting Gaussian beams. The conditions for the obtained diffraction-free wave to be physically meaningful are discussed.
\end{abstract}

\maketitle

\section{Introduction}

\noindent Recently, space-time beams \cite{KONDAKCI}, needle pulses \cite{ALONSO} and abruptly focusing needles of light \cite{KAMINER} are arousing considerable interest as an alternate way to eliminate or reduce diffraction spreading of light beams. Instead of beam-shaping approaches, as in Airy or Bessel beams \cite{SIVILOGLOU,DURNIN}, the underlying idea in \cite{KONDAKCI,ALONSO,KAMINER} is, for a beam propagating mainly in the $z$ direction, to get rid of the monochromatic constrain and to force all monochromatic plane-wave (MPW) spectral constituents, $(\mathbf{k}_\perp,k_z,\omega)$, $\mathbf{k}_\perp=(k_x,k_y)$, linked by $(|\mathbf{k}_\perp|^2+k_z^2)^{1/2}=\omega/c$, to have the same $k_z$ component. Allowing for a tolerance in $k_z$, the pulsed beam carries finite energy, forming an arbitrarily long, but not infinite, needle of light. Seemingly contradictory values for the group velocity of these propagation-invariant fields are awarded, e. g., zero in \cite{ALONSO}, $c$ (the speed of light in vacuum) in \cite{KAMINER}, or infinite, as seen here. Refs. \cite{ALONSO} and \cite{KAMINER} provide closed-form expressions, free of approximations, for these beams, valid therefore for extreme focusing of few-cycle pulses. The analogy between the expression of a monochromatic light beam, $A(\mathbf{r}_\perp,z)e^{-i\omega t}$, and $A(\mathbf{r}_\perp,t)e^{ik_z z}$ for these nondiffracting pulsed beams led to the concept of {\it temporal diffraction}, and to call them time-diffracting (TD) beams \cite{KAMINER}. The analogy is, however, limited, since the wave equation, $\Delta E- (1/c^2)\partial_{tt}E=0$, with the two above ansatzs does not lead to formally equal Helmholtz equations \cite{NOTE}.

As pointed out earlier \cite{SAARI,PORRAS}, the general condition of propagation-invariance is $k_z=a+\omega/v$, $v$ being the group velocity. The different families of propagation-invariant {\it localized waves} (LWs) are considered and classified in \cite{SAARI} for free-space propagation, and in \cite{PORRAS,PORRASOL} in dispersive media. On the other hand, simple models as the Schr\"odinger equation for paraxial propagation of quasi-monochromatic (QM) light, and the Gaussian beam (GB) solution in particular, represent scientific milestones, even if they are only approximate, to which any advance in light beam propagation makes reference. Paraxial models also ease the study of nonlinear propagation phenomena, commonly modeled by nonlinear Schr\"odinger equations, and the transformation by optical systems.

Here we describe closed-form solutions of the paraxial wave equation that describe diffraction-free pulsed beams travelling at rather arbitrary speeds. Diffraction is explicitly shown to develop in time. We make emphasis in TD Gaussian beams (TDGBs), take special care in delimiting the conditions for the paraxial TD beams to be physically meaningful, and relate them to previously known LW families. Finite-energy, quasi-nondiffracting TD solutions are also obtained as TD beams enveloped by a luminal plane pulse. The particular TDGBs with infinite speed are seen to be paraxial, many-cycle cases of the light needles in \cite{KONDAKCI,ALONSO,KAMINER}. From the simplicity of our approach, we clarify the issue of the propagation speed.

\section{Time-diffracting Gaussian beams at arbitrary speeds}

For the wave packet $E=A(\mathbf{r}_\perp,z,t)e^{-i(\omega_0 t - k_0 z)}$ of carrier frequency $\omega_0$ and propagation constant $k_0=\omega_0/c$, the wave equation yields, in terms of the variables $t'=t-z/c$ and $z'=z$, $\Delta_\perp A +\partial_{z'z'} A+ 2ik_0 \partial_{z'} A =(2/c)\partial_{z't'} A$,
%
%
with $\Delta_\perp=\partial_{xx}+\partial_{yy}$. Paraxial propagation allows to neglect the second derivative in $z'$. Spatiotemporal (ST) coupling effects {\it arising from propagation} are accounted for by the cross derivative \cite{PORRAS2}. We can also neglect these effects by considering many-cycle, QM pulses (e. g., a few tenths of femtoseconds or longer for visible light) \cite{PORRAS2,PORRAS3}. These couplings are also very small compared to the intentionally introduced ST couplings of TD beams. Paraxial propagation of QM pulses can then be described by
\begin{equation}\label{PAR}
\Delta_\perp A +2ik_0 \partial_{z'} A=0 ,
\end{equation}
i. e., by the usual paraxial wave equation, in which time does not appear explicitly. Since (\ref{PAR}) admits the factorized solutions in time and space, paraxial propagation is commonly described as an undistorted pulse at speed $c$ except for an amplitude change from point to point due to paraxial (Fresnel) diffraction. However, the factorized form is only one possibility.

We search for physically valid, nonseparable solutions of the form $A=A_\alpha(\mathbf{r}_\perp,t'+\alpha z')$, which represent nondiffracting pulsed beam propagation at superluminal or subluminal velocity $v=c/(1-\alpha c)$. With this anzatz, (\ref{PAR}) yields
\begin{equation}\label{PART}
\Delta_\perp A_\alpha +2ik_0\alpha \partial_{t'} A_\alpha=0
\end{equation}
for the spatiotemporal shape of $A_\alpha(\mathbf{r}_\perp,t')$. Equation (\ref{PART}) is the same as (\ref{PAR}) except that diffraction appears to take place in time. Thus, if we take any of the known solutions $A(\mathbf{r}_\perp,z')$ of (\ref{PAR}) representing a (strictly) monochromatic light beam, replace $z'$ with $t'+\alpha z'$, $k_0$ with $k_0|\alpha|$, and $A$ with $A^\star$ in case that $\alpha<0$, we obtain a nondiffracting pulsed beam satisfying also (\ref{PAR}). The solutions constructed in this way will be called here TD beams. Transversal localization of the original monochromatic light beam ensures transversal localization of the TD beam. Localization of the original beam caused by diffraction ensures temporal localization of the TD beam. Given the prominent role of GBs in light beam propagation, we focus most of our attention in its TD counterpart, or TD Gaussian beam (TDGB), given by
\begin{equation}\label{TDGB}
A_\alpha =\frac{-it_0}{p(t')}\exp\left[\frac{ik_0|\alpha| r^2}{2p(t')}\right] ,
\end{equation}
where $r=|\mathbf{r}_\perp|$, $p(t')= t'+\alpha z' -i t_0$ and $t_0>0$, in case that $\alpha>0$, and the complex conjugate of the rhs of (\ref{TDGB}) in case that $\alpha<0$. The peak intensity is assumed to be one in adequate units. For the luminal velocity $v=c$ ($\alpha=0$), (\ref{TDGB}) is a plane pulse of envelope $A_0=-it_0/p(t')$, whose intensity FWHM duration is $\Delta t= 2t_0$. For other velocities ($\alpha\neq 0$), $-it_0/p(t')$ is also the on-axis pulse shape, but the radial amplitude is Gaussian function $\exp[-r^2/w^2(t')]$ of width $w^2(t')=(2t_0/k_0|\alpha|)[1+(t'+\alpha z')^2/t_0^2]$ changing hyperbolically in time, as the width of the monochromatic GB does in $z$. The waist width is then $w_0=\sqrt{\Delta t/k_0|\alpha|}$ at the pulse center $t'+\alpha z'=0$.

Some restrictions should be set on the parameters $t_0$ and $\alpha$ for (\ref{TDGB}) to be physically meaningful. Requiring the bandwidth, which can be characterized by $\Delta \omega=1/t_0$, to be much smaller than $\omega_0$, and the transversal size $w_0$ to be significantly larger than the wave length $\lambda_0=2\pi/k_0$, we set
\begin{equation}\label{COND}
\frac{1}{t_0}=\Delta \omega \ll \omega_0, \quad |\alpha| \ll \frac{\omega_0 t_0}{2\pi^2 c}=\frac{\omega_0}{2\pi^2 c\Delta \omega} .
\end{equation}

Figures \ref{Fig1} (a) and (b) show the ST intensity distribution and the temporal pulse shapes at selected radii for a particular TDGB satisfying conditions (\ref{COND}). One of the most accessible properties of a pulsed beam in experiments is the transversal fluence profile  ${\cal F}(\mathbf{r}_\perp,z)=\int |A_\alpha|^2 dt'$. For TD beams, it is independent of $z'$, and for the TDGB it is found to be given (see the spectral analysis) by ${\cal F}(r)= \pi t_0 I_0(r^2/w_0^2)\exp(-r^2/w_0^2)$, where $I_0(\cdot)$ is the modified Bessel function of the first kind and order zero. An example is plotted in Fig. \ref{Fig1}(c).

\subsection*{Finite-energy realizations}

As all nondiffracting beams described so far, TD beams involve some kind of idealization: Since the power of the original monochromatic light beam is independent of axial distance, the instantaneous power $P(t')=\int |A_\alpha|^2 d\mathbf{r}_\perp$ of the TD counterpart is finite but independent of time, and therefore the TD energy ${\cal E}= \int P(t') dt'$ is infinite. For the TDGB, the instantaneous power is $P(t')=\pi w_0^2/2$, and hence ${\cal E}=\infty$. This can also be seen from the asymptotic decay ${\cal F}(r)\sim 1/r$ of the fluence profile [$I_0(s)\sim e^s/(2\pi s)^{1/2}$ at large $s$], which yields infinite energy upon integration in the transversal plane.

\begin{figure}[!]
\centering
\includegraphics*[height=4.2cm]{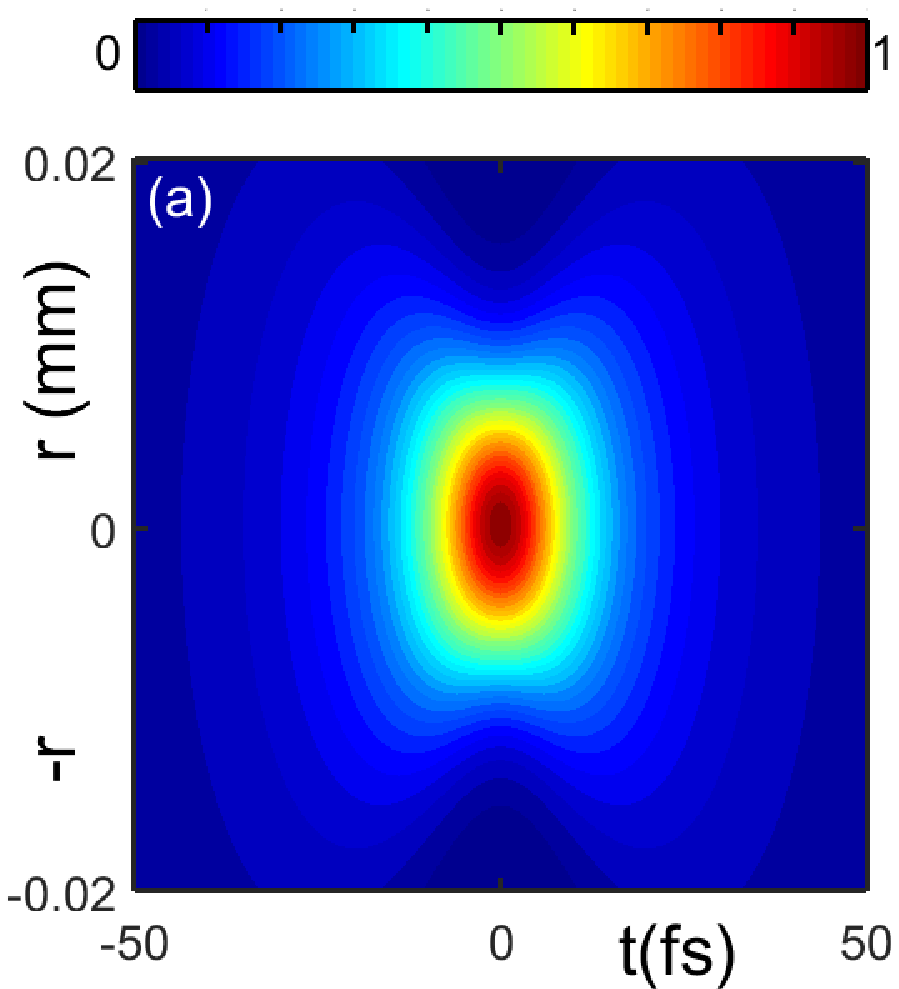}\includegraphics*[height=4.2cm]{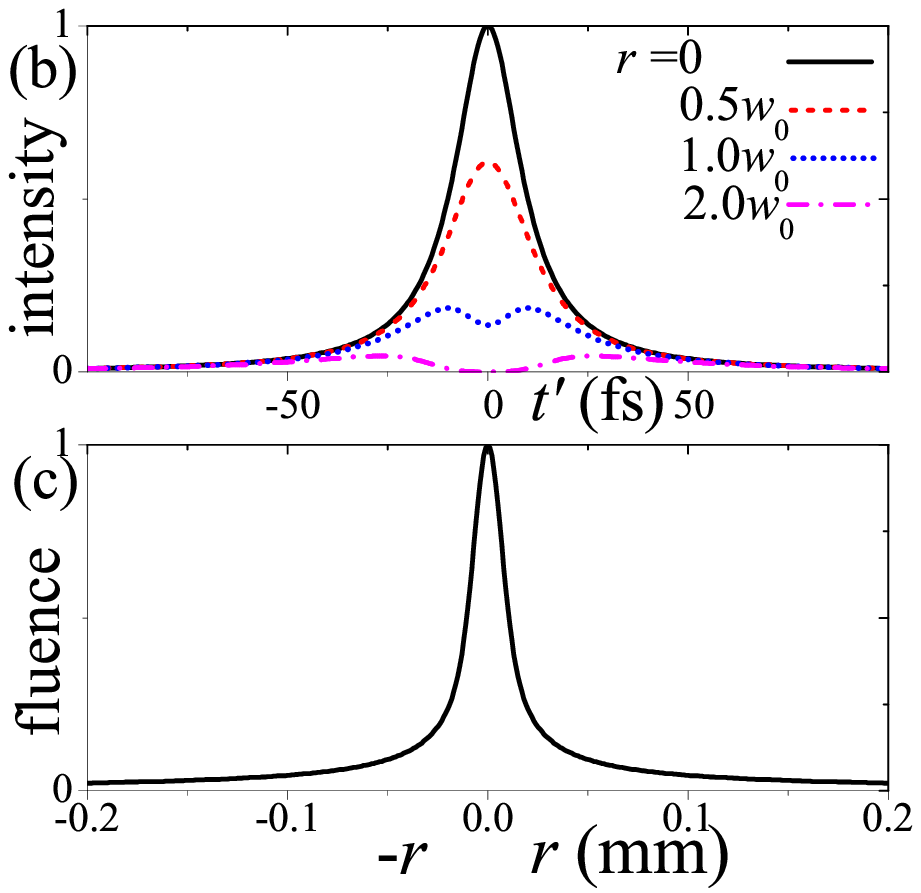}
\caption{\label{Fig1} TDGB of carrier wave length $\lambda_0=2\pi/k_0=800$ nm, with $t_0=10$ fs (FWHM $\Delta t=20$ fs) and $\alpha=20$ fs/mm. With this choice the velocity is $v=3.018\times 10^{-4}$ mm/fs, slightly superluminal ($c=3\times 10^{-4}$ mm/fs), and QM and paraxial conditions, $\Delta \omega=0.1\ll \omega_0= 2.356$ fs$^{-1}$ and $\alpha < \omega_0 t_0/2\pi^2 c=3979$ fs/mm, are well satisfied. (a) ST intensity distribution at $z'=0$. (b) Pulse shapes at radii $r=0,w_0/2,w_0$ and $2w_0$, with $w_0=0.011$ mm, at $z'=0$. (c) Radial fluence profile. These properties are independent of $z'$, except a time shift $-\alpha z'$ in the local time $t'$ due to the superluminal velocity.}
\end{figure}

Finite-energy versions of the above TD beams approximating the ideal behavior for arbitrarily long distances can be written down straightforwardly. We note that $A(\mathbf{r}_\perp,t',z')=f(t')A_\alpha(\mathbf{r}_\perp,t'+\alpha z')$, where $f(t')$ represents a luminal plane pulse envelope, still satisfies the paraxial wave equation (\ref{PAR}) if $A_\alpha(\mathbf{r}_\perp,t')$ satisfies (\ref{PART}). In particular,
\begin{equation}\label{TDGBF}
A(r,t',z')=f(t')\frac{-it_0}{p(t')}\exp\left[\frac{ik_0|\alpha| r^2}{2p(t')}\right]
\end{equation}
satisfies (\ref{PAR}). The instantaneous power of (\ref{TDGBF}) is now $P(t')=(\pi w_0^2/2)|f(t')|^2$, and then its energy ${\cal E}=(\pi w_0^2/2)\int_{-\infty}^\infty|f(t')|^2 dt'$ is finite with standard choices of pulse shapes such as $f(t')=\exp(-t^{\prime 2}/\tau^2)$. The situation of interest is that in which $f(t')$ is a smooth pulse much longer than $A_\alpha$. The product $f(t')A_\alpha$ behaves approximately as the ideal TDGB while $f(t')$ and $A_\alpha$ overlap. If $\Delta t_f$ is the intensity FWHM duration of $f(t')$, $A_\alpha$ crosses $f(t')$ in a distance $L_{\rm free}$ given by $|\alpha|L_{\rm free}=\Delta t_f$, i. e., $L_{\rm free}= \Delta t_f/|\alpha|$. In units of the confocal parameter (twice the Rayleigh distance) $L_{\rm R}=2z_{\rm R}=k_0 w_0^2=\Delta t/|\alpha|$ associated with the monochromatic GB of the same wave length and waist width, the diffraction-free distance is $L_{\rm free}/L_{\rm R}=\Delta t_f/\Delta t$.

\begin{figure}[t!]
\centering
\includegraphics*[height=4cm]{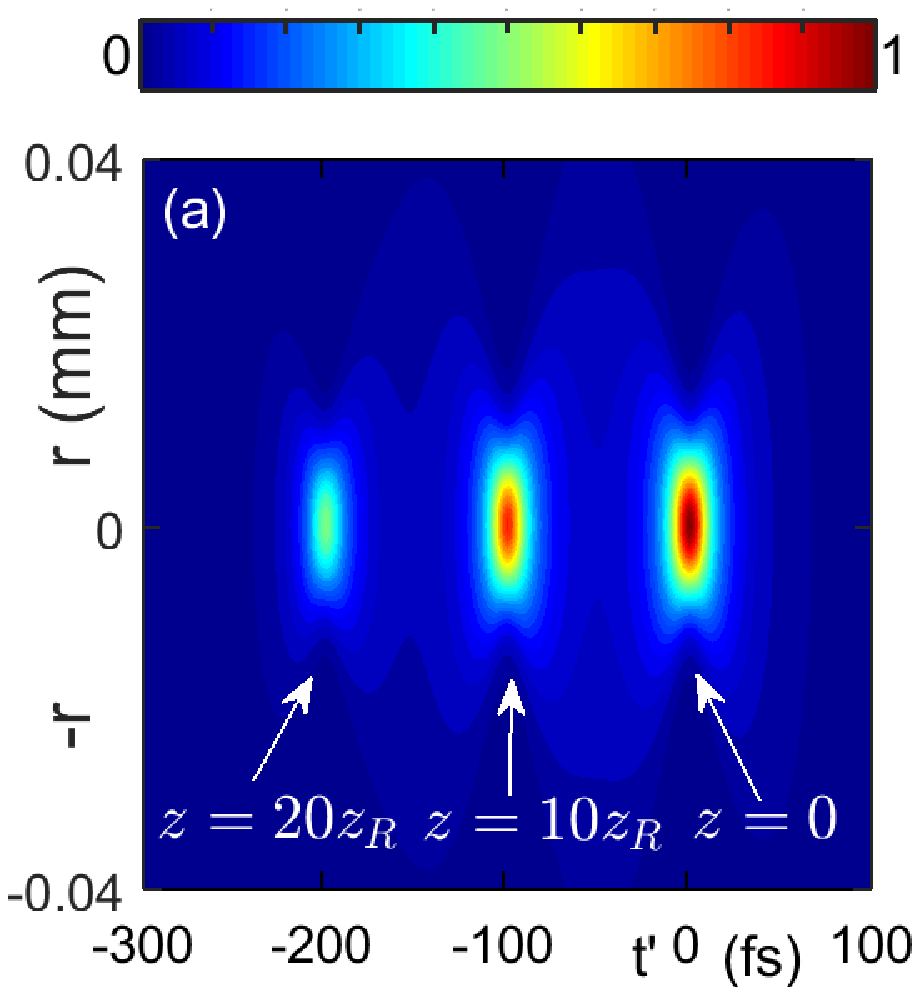}\includegraphics*[height=4cm]{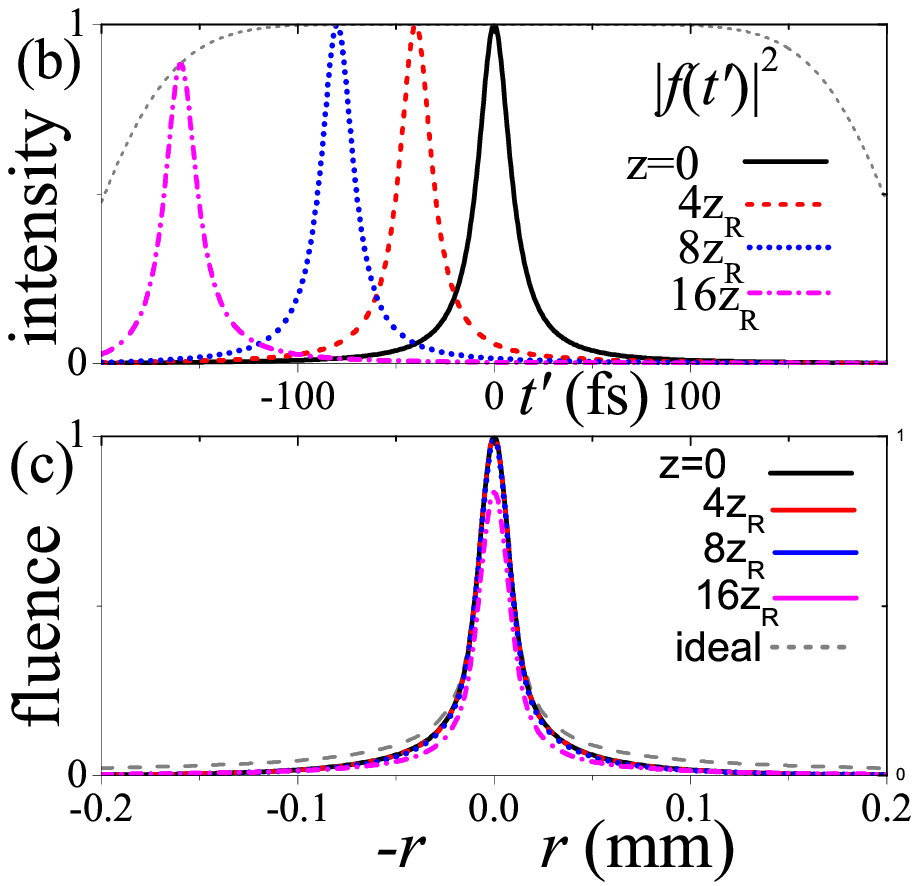}
\includegraphics*[width=7.9cm]{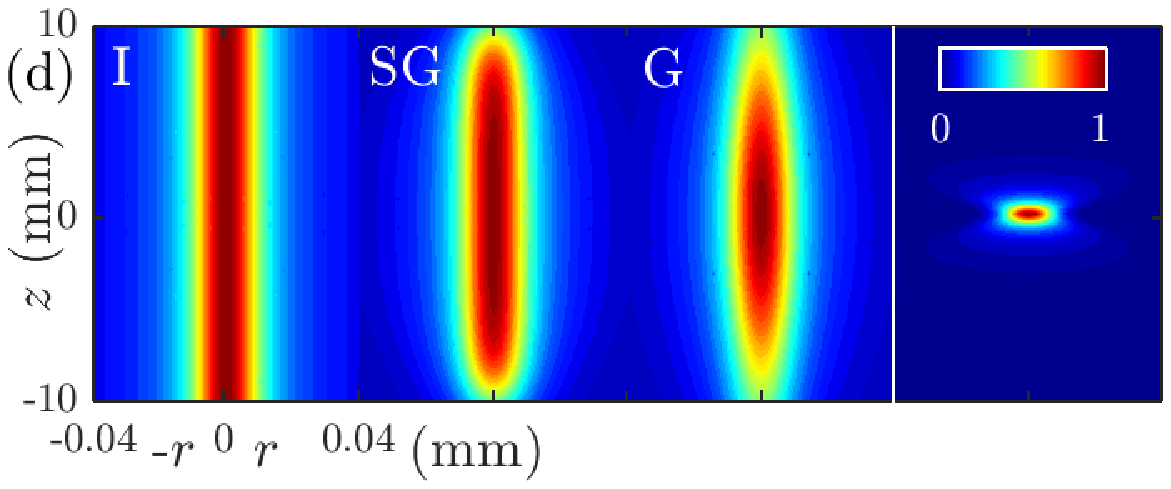}
\caption{\label{Fig2} (a-c) Finite-energy TDGB at $\lambda=800$ nm, with $t_0=10$ fs, $\alpha=20$ fs/mm, and $f(t')=\exp(-t^{\prime 8}/\tau^8)$, $\tau=226.5$ fs, or FWHM duration $\Delta t_f=400$ fs. The diffraction-free length is $L_{\rm free}=20$ mm (from $-10$ to $+10$ mm), which is $20$ times the confocal length $L_R=2z_R=1$ mm (from $-0.5$ to $+0.5$ mm) for the given waist width. (a) ST profiles, (b) on-axis pulse shapes and (c) fluence profiles at the indicated propagation distances.
(d) Fluence profiles of the ideal TDGB (I), of the finite-energy TDGB with the above super-Gaussian $f(t')$ (SG), and with Gaussian $f(t')=\exp(-t^{\prime 2}/\tau^2)$, $\tau=339.7$ fs ($\Delta t_f=400$ fs too). In the right, intensity profile of a monochromatic GB of the same wave length and waist width.}
\end{figure}

For the TDGB in Fig. \ref{Fig1} but with finite energy, Fig. \ref{Fig2}(a-c) shows the quasi-invariant ST intensity distributions, on-axis pulse shapes and fluence profiles at selected propagation distances for super-Gaussian $f(t')$ of duration such that $L_{\rm free}$ is $20$ times $L_{\rm R}$. The fluence profiles [solid curves in Fig. \ref{Fig2} (c)] are more localized than that of the ideal TDGB (dashed curve). The fluence profiles of the ideal TDGB, of the above finite-energy TDGB, and a finite-energy TDGB with Gaussian $f(t')$ (also such that $L_{\rm free}=20 L_R$), and the intensity profile of a monochromatic GB of the same waist width, are plotted in Fig. \ref{Fig2} (d) to evidence the diffraction-free propagation along 20 times $L_{\rm R}$. With equal radial and axial scales, the fluences of the finite-energy TDGBs will appear as extremely narrow, $2w_0=22\,\mu$m wide, $20$ mm long, needles of light.

\section{Spectral analysis}

A spectral analysis provides another insight into the nondiffracting behavior of TD beams. Following the space-time analogy, the solution of (\ref{PART}) with a given spectrum $\hat g(\mathbf{k}_\perp)$ of transversal frequencies is
\begin{equation}\label{TDS}
A_\alpha(\mathbf{r}_\perp,t'\!+\!\alpha z')\!=\!\frac{1}{2\pi}\int \!d\mathbf{k}_\perp \hat g(\mathbf{k}_\perp) e^{-i\frac{|\mathbf{k}_\perp|^2}{2k_0\alpha}(t'+\alpha z')} e^{i\mathbf{k}_\perp\cdot\mathbf{r}_\perp}.
\end{equation}
We can then conveniently write the complete electric field as
\begin{equation}
E(\mathbf{r}_\perp,z,t)=\frac{1}{2\pi} \int d\mathbf{k}_\perp \hat g(\mathbf{k}_\perp) e^{i \mathbf{k}_\perp\cdot\mathbf{r}_\perp} e^{ik_z(\Omega) z} e^{-i(\omega_0+\Omega) t},
\end{equation}
where $\Omega=|\mathbf{k}_\perp|^2/2k_0\alpha$, and $k_z(\Omega)=k_0+ [(1/c)-\alpha]\Omega = k_0 + \Omega/v$.
We observe that the temporal frequency of each MPW constituent of transversal wave vector $\mathbf{k}_\perp$ is shifted from $\omega_0$ by $\Omega$ in such a way that the axial component varies linearly with frequency. Transversal components of the wave vectors and temporal frequencies are then linked in TD beams by
\begin{equation}\label{Kp}
|\mathbf{k}_\perp|^2=2k_0\alpha\Omega.
\end{equation}
Since $k_z$ cannot be larger than $k=\omega/c$ ($|\mathbf{k}_\perp|^2$ cannot be negative), temporal frequencies in TD beams are limited to $\Omega>0$ for $\alpha>0$, and to $\Omega<0$ for $\alpha<0$. Eq. (\ref{Kp}) is the parabolic approximation about $\Omega=0$ to the exact (hyperbolic, parabolic or elliptic) relation $|\mathbf{k}_{\perp,\rm exact}|^2=[2k_0 - (\alpha-2/c)\Omega]\alpha\Omega$, obtained using $k_z=k_0+[(1/c)-\alpha]\Omega$ in the dispersion relation $|\mathbf{k}_\perp|^2 + k_z^2=(\omega_0+\Omega)^2/c^2$ of the wave equation, or light cone. Conditions (\ref{COND}) are readily seen to imply that the terms with $\Omega^2$ in $|\mathbf{k}_{\perp,\rm exact}|^2$ are negligibly small compared to the term with $\Omega$.

From (\ref{TDS}), finite-energy TD beams, $A(\mathbf{r}_\perp,t',z')= f(t')A_\alpha(\mathbf{r}_\perp,t'+\alpha z')$, or ideal TD beams [$f(t')=1$] can be written as
\begin{equation}
A(\mathbf{r}_\perp,t',z')=\frac{1}{2\pi}\int \!d\Omega\int \!d \mathbf{k}_\perp \hat A(\mathbf{k}_\perp,\Omega,z')e^{-i\Omega t'}
e^{i\mathbf{k}_\perp\cdot \mathbf{r}_\perp},
\end{equation}
where the ST spectrum is
\begin{equation}\label{SPF}
\hat A (\mathbf{k}_\perp,\Omega,z')=\frac{1}{2\pi}\hat g(\mathbf{k}_\perp)\hat f\left(\Omega-\frac{|\mathbf{k}_\perp|^2}{2k_0\alpha}\right)
e^{-i\frac{|\mathbf{k}_\perp|^2}{2k_0} z'},
\end{equation}
$\hat f(\Omega)$ is the Fourier transform of $f(t')$, and $\hat f(\Omega-|\mathbf{k}_\perp|^2/2k_0\alpha)$ is to be replaced with $2\pi \delta(\Omega -|\mathbf{k}_\perp|^2/2k_0\alpha)$ for ideal TD beams. If $f(t')$ is a long pulse, the ST spectrum is highly concentrated about the curve $|\mathbf{k}_\perp|^2=2k_0\alpha\Omega$. This imposes also the correlation $\Delta k_\perp^2=k_0|\alpha|\Delta \omega$ between the spatial and temporal bandwidths in TD beams. The choice of the Gaussian transversal spectrum $\hat g(\mathbf{k}_\perp)= (t_0/k_0|\alpha|)\exp[-t_0^2|\mathbf{k}_\perp|^2/2k_0|\alpha|]$ of width $\Delta k_\perp =2/w_0$, (\ref{TDS}) yields the TDGB in (\ref{TDGB}). The amplitude of each frequency $\Omega=|\mathbf{k}_\perp|^2/2k_0\alpha$, or temporal spectrum of the TDGB, is then given by $\hat g(\Omega)=(t_0/k_0|\alpha|)\exp(-t_0|\Omega|)$, of bandwidth $\Delta \omega=1/t_0$. The relation $w_0^2=\Delta t/k_0|\alpha|$ for TDGBs is the particular expression of $\Delta k_\perp^2=k_0|\alpha|\Delta \omega$ for general TD beams.

\begin{figure}[t!]
\centering
\includegraphics*[width=4cm]{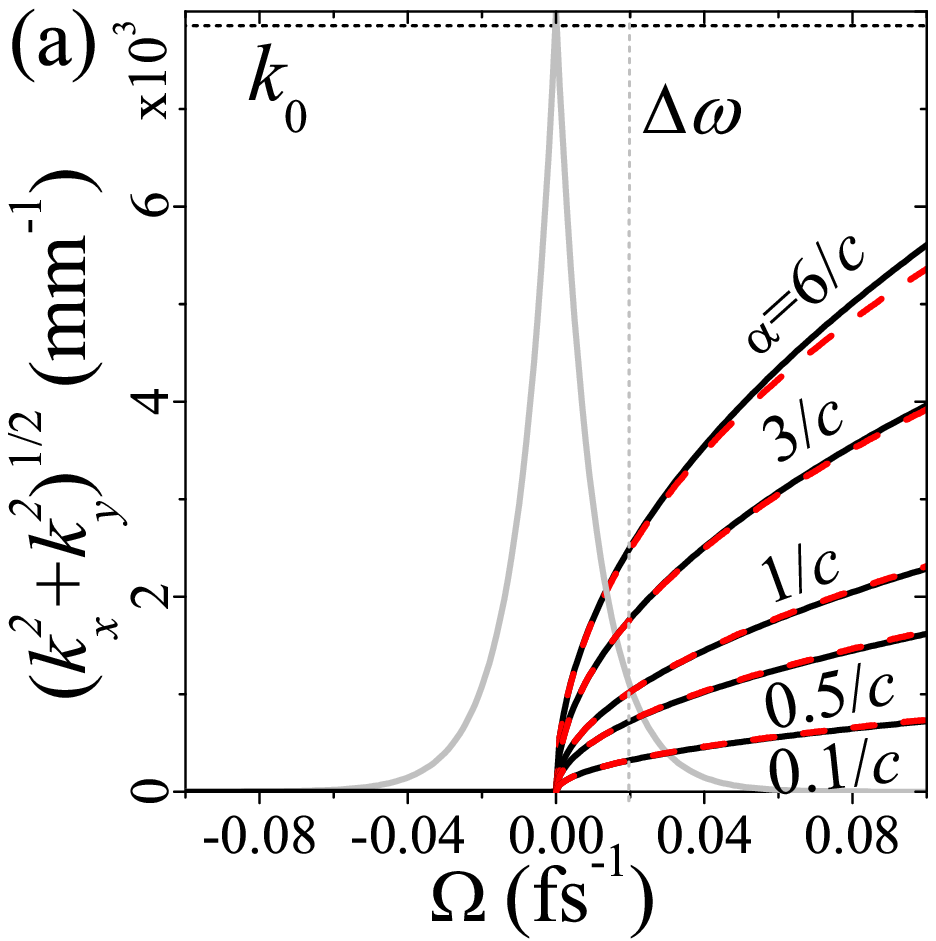}\includegraphics*[width=4cm]{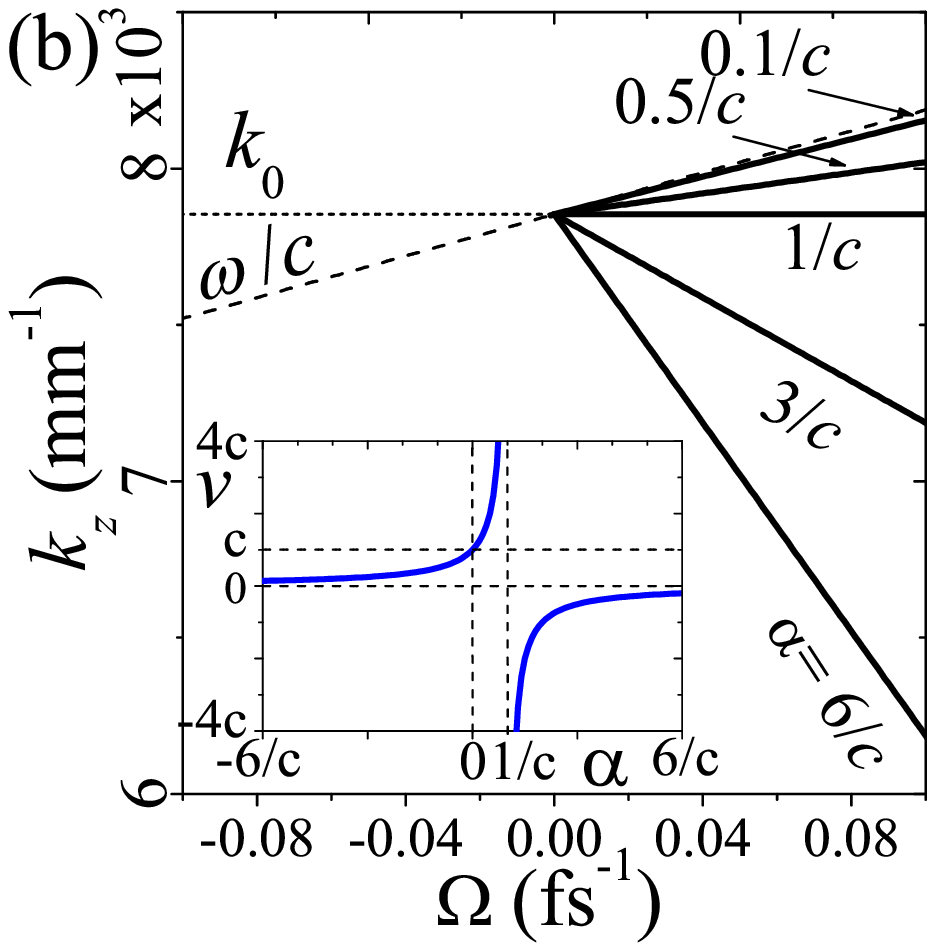}
\includegraphics*[width=4cm]{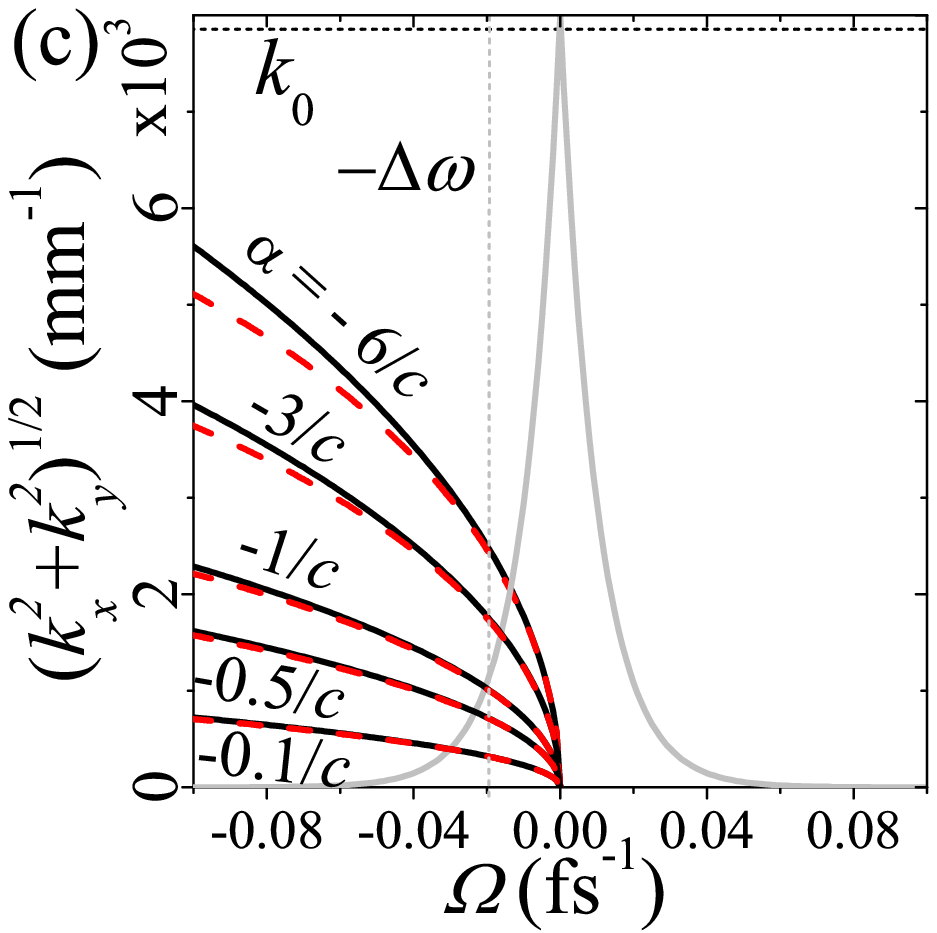}\includegraphics*[width=4cm]{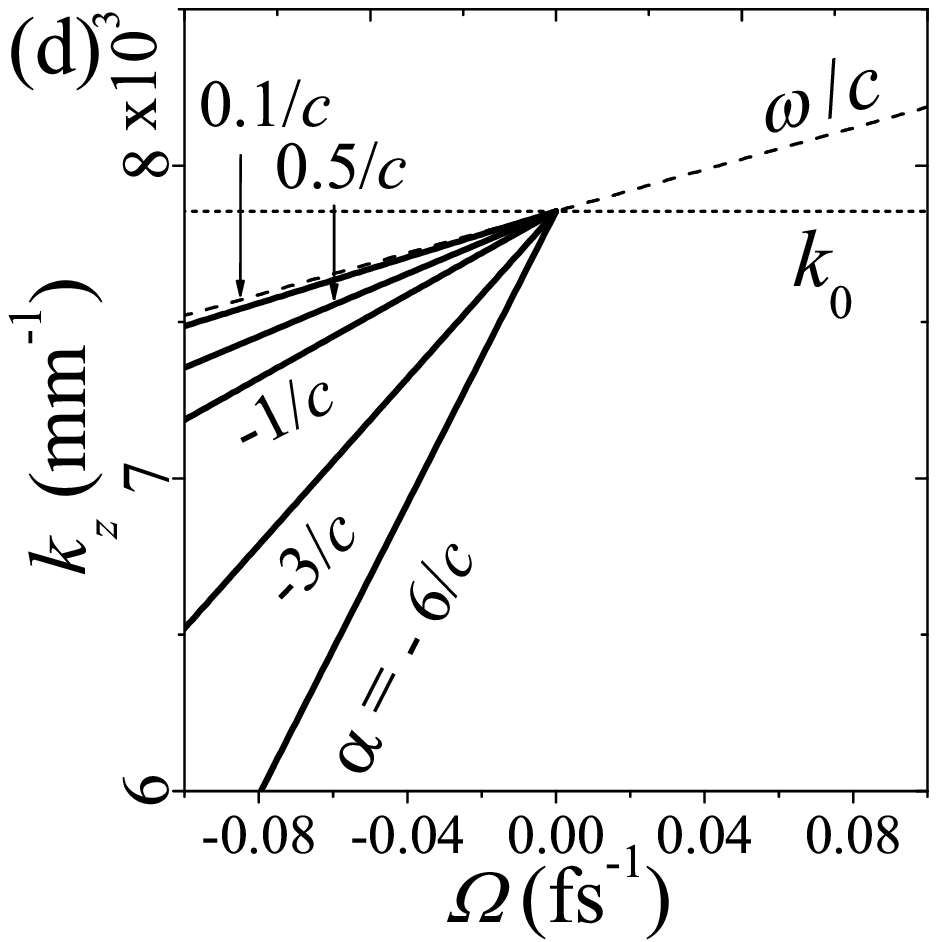}
\caption{\label{Fig3} For TDGBs with $\lambda_0=800$ nm, $t=50$ fs and the indicated positive values of $\alpha$, (a) modulus of the transversal component of the wave vector, evaluated from (\ref{Kp}) (solid curves) and from the exact relation $\kappa_{\rm NP}(\Omega)$ (dashed red curves), and (b) longitudinal component (solid curves), as functions of frequency. (c) and (d) The same for the indicated negative values of $\alpha$. For reference, the gray curves represent the temporal spectral density, in arbitrary units, of the TGDB of the bandwidth $\Delta\omega=1/t_0$. The inset in (b) depicts the propagation velocity $v$ as a function of $\alpha$.}
\end{figure}

Figure \ref{Fig3} illustrates the above spectral properties. As a reference, the inset in Fig. \ref{Fig3}(b) shows the velocity of TD beams of duration $t_0=50$ fs at $\lambda_0=800$ nm for the range of values of $\alpha$ that comply the condition of paraxiality, $|\alpha|\ll \omega_0 t_0/2\pi^2c$, or $|\alpha|\ll 6/c$ in this particular example. As seen, subluminal, luminal and superluminal velocities, both positive and negative, are included in this range. For a few allowed values of $\alpha$, Figs. \ref{Fig3}(a) and (b) (for $\alpha>0$) and \ref{Fig3} (c) and (d) ($\alpha<0$) depict $|\mathbf{k}_\perp|$ and $k_z$ at each frequency $\Omega$. For reference, we also plot the temporal spectral density $|\hat g(\Omega)|^2$ of bandwidth $\Delta\omega=1/t_0$ (gray curves), $k_0$ (horizontal dotted lines), and the modulus of the wave vector $\omega/c$ (dashed lines). As seen, $|\mathbf{k}_\perp|^2$ evaluated from the paraxial relation (\ref{Kp}) and from the exact relation $|\mathbf{k}_{\perp,\rm exact}|^2$ [solid and dashed curves in Figs. \ref{Fig3} (a) and (c)], are almost indistinguishable and paraxial (much smaller than $k_0$) {\it within the bandwidth}. For $\alpha=1/c$, corresponding to $v=\infty$, the most inclined MPW constituent within the bandwidth travels as $10^\circ$ from the beam axis,
thus supporting the validity of the paraxial TDGB in (\ref{TDGB}) to describe the nondiffracting propagation at these velocities.

\begin{figure}[t!]
\centering
\parbox[b]{4.5cm}{
\includegraphics*[width=4.3cm]{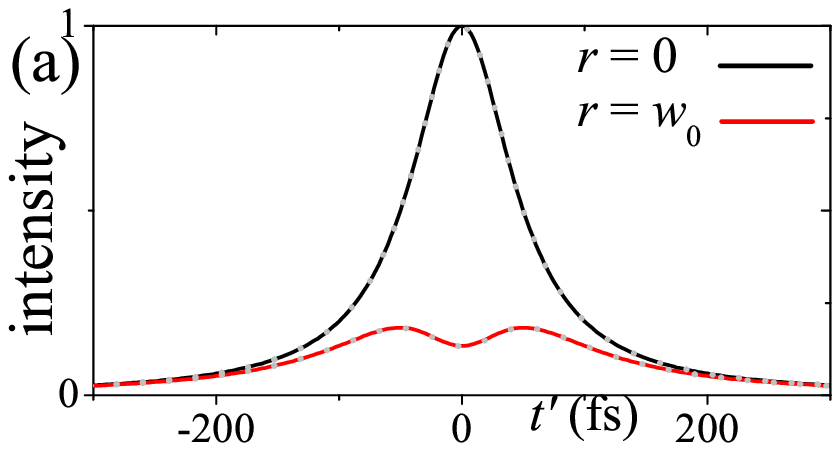}
\vspace{0.2cm}
\includegraphics*[width=4.5cm]{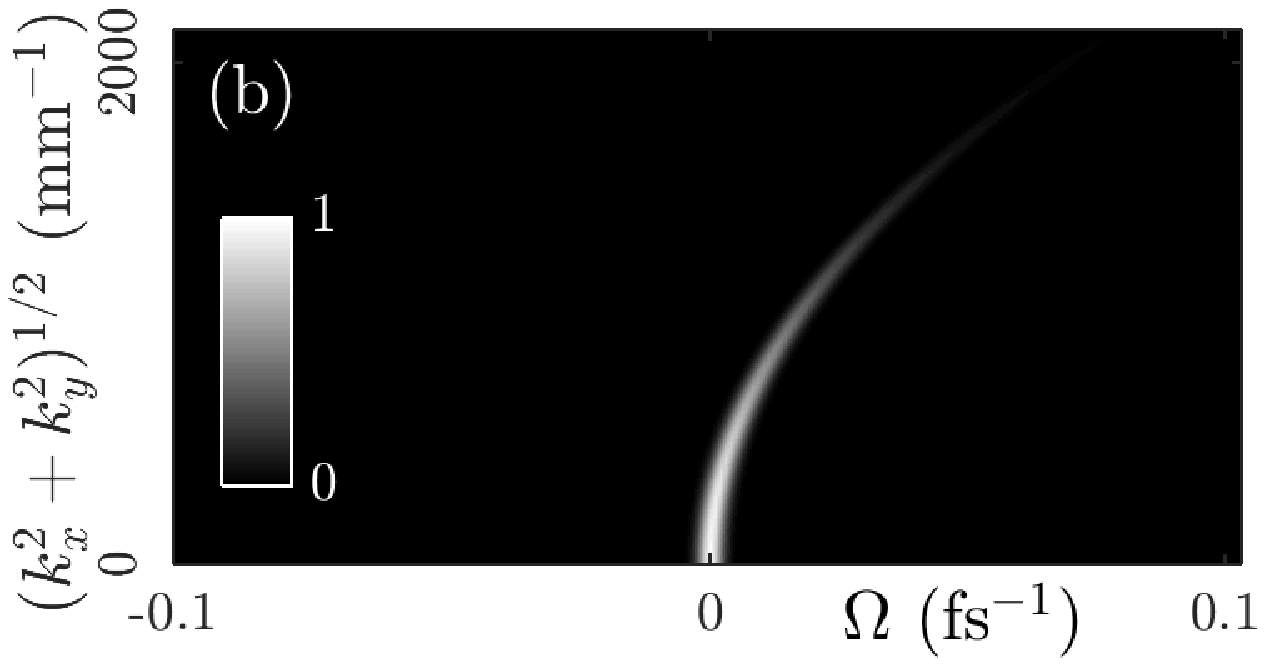}}
\parbox[t]{3.8cm}{
\includegraphics*[width=3.8cm]{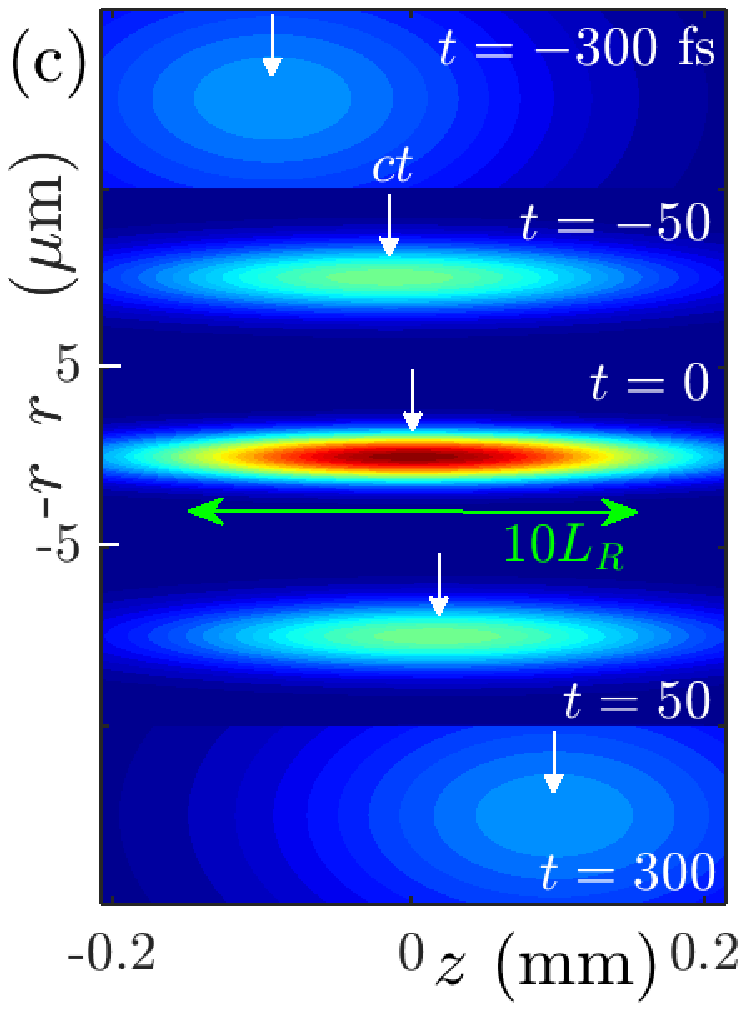}}
\caption{\label{Fig4} (a) For the ideal TDGB at $800$ nm, $\Delta t=2t_0=100$ fs, and $\alpha=1/c$, pulse shapes at radii $r=0$ and $r=w_0\simeq 2\,\mu$m and $z'=0$ according to (\ref{TDGB}) (solid curves) and to Eq. (18) in \cite{ALONSO} (dashed curves). (b) Spectral density $|\hat A|^2$ given by (\ref{SPF}) of the finite-energy realization with $f(t')=\exp(-t^{\prime 2}/\tau^2)$, with $\tau=850$ fs ($\Delta t_f=1000$ fs). (c) Snapshots of the spatial intensity distribution, $|A|^2$, at the indicated instants of time $t$.}
\end{figure}

\section{Discussion}

The TDGBs with $\alpha=1/c$ ($v=\infty$, $k_z=k_0=\mbox{const.}$) are QM and paraxial cases of the needles of light, or TD beams reported in \cite{KONDAKCI,ALONSO,KAMINER}. For example, the needle pulse in Eq. (18) in \cite{ALONSO} is actually the cylindrically converging-expanding wave reported in \cite{SAARI}, and reduces (with the identifications $k_L=k_0$ and $q=t_0$) to the ideal TDGB with $\alpha=1/c$  when $r\sim w_0\ll ct_0$, which is the same as condition $1/t_0\ll \omega_0$ in (\ref{COND}). For the same duration as in Fig. \ref{Fig3}, Fig. \ref{Fig4}(a) shows no difference between the TDGB (\ref{TDGB}) and Eq. (18) in \cite{ALONSO}, even if the waist diameter is only $4\, \mu$m. Also, Eqs. (8) and (9) in \cite{KAMINER} for TD beams, and Eq. (10) for the relation between temporal and transversal extents, are the electric field of the finite-energy TDGB with $\alpha=1/c$, Gaussian $f(t')$, and our relation $\Delta t=k_0|\alpha| w_0^2$ with  $\alpha=1/c$. We plot in Fig. \ref{Fig4}(b) the ST spectral density $|\hat A|^2$ of a finite-energy realization with $f(t')=\exp(-t^{\prime 2}/\tau^2)$ ($\tau=850$ fs, or $\Delta t_f=1000$ fs) such that $L_{\rm free}=\Delta t_f/|\alpha|=c\Delta t_f$ is $10$ times $L_{\rm R}=\Delta t/|\alpha|=c\Delta t$. The temporal dynamics of the TDGB with $v=\infty$, as viewed from the local time frame $t'$, does not qualitatively differ from what is shown in Fig. \ref{Fig2} for other velocities, but things look quite different from the laboratory time $t$. For example, the time shift of the curves in Fig. \ref{Fig2}(b) would now be $-z/c$ for the respective values of $z$, meaning that the pulse arrives at the same time $t$ at any distance $z$. This is because, although $\Delta t=2t_0$ is much smaller that $\Delta t_f$, the axial length, $v \Delta t$, of the ideal TDGB is larger, infinitely larger indeed, that the axial length, $c \Delta t_f$, of the luminal pulse $f(t-z/c)$, a length that coincides with $L_{\rm free}$. All together, the finite-energy TDGB with $v=\infty$ is a narrow needle of light of length $L_{\rm free}=10 L_{\rm R}$ that appears and disappears simultaneously at all distances in a lapse of time $\Delta t$, as illustrated by the three snapshots of Fig. \ref{Fig4}(c). In this short lapse of time, the needle of light delimited by $f(t-z/c)$, advances the distance $c\Delta t$, as indicated by the white arrows in the center of $f(t-z/c)$, which makes the effective velocity to be equal to $c$, as reported in \cite{KAMINER}. We focused here on the TDGB with $\alpha=1/c$ for comparison purposes, but similar needles of light are formed, as in Fig. \ref{Fig2}, with other superluminal, even negative velocities, except that the TDGB pulse (no longer of infinite length) is "seen" moving forwards or backwards along the envelope.

TD beams are obviously particular LWs \cite{SAARI} ($k_z=a+\omega/v$) with the peculiarities that 1) $k_z= k_0+(\omega -\omega_0)/v$ intersects the light cone $k=\omega/c$ at a positive frequency $\omega_0$, or equivalently, $|\mathbf{k}_{\perp,\rm exact}|$ is a hyperbola, parabola or ellipse crossing the $\omega$-axis at $\omega_0$ ($\Omega=0$), and 2) temporal frequencies are close enough to $\omega_0$ for the parabola $|\mathbf{k}_\perp|=(2k_0\alpha\Omega)^{1/2}$ to approximate $|\mathbf{k}_{\perp,\rm exact}|$. TD beams thus exist belonging to all families of LWs, as classified in \cite{SAARI}, except to the Bessel-X type (the crossing point is $\omega=0$), e. g., TD beams with $|v|>c$ ($0<\alpha<2/c$) belong to the superluminal LW family ("focused X waves"), and with $|v|<c$ ($\alpha<0$ and $\alpha>2/c$) to the subluminal LW family. As said, the luminal TD beam with $v=c$ is a trivial plane pulse, but the luminal TD beam with $v=-c$ ($\alpha=2/c$) is particularly intriguing: $|\mathbf{k}_{\perp,\rm exact}|$ is the same parabola as $|\mathbf{k}_\perp|=(2k_0\alpha\Omega)^{1/2}$, and the TD beam is the focus wave mode $E=A_{2/c}(\mathbf{r}_\perp,t'+2z/c=t+z/c)e^{-i(\omega_0 t-k_0 z)}$, which is known to be an {\it exact} solution of the wave equation (not restricted therefore to paraxial and QM conditions) if $A_{2/c}(\mathbf{r}_\perp,t')$ satisfies the paraxial wave equation in time (\ref{PART}) \cite{SEZGINER}. In this connection, we have generalized this fact to arbitrary speeds, with the difference that the obtained TD beam is not an exact but approximate (paraxial) solution to the wave equation.

\section{Conclusion}

This work shows, to conclude, that the simple Schr\"odinger equation ruling paraxial wave optics has not yet been sufficiently explored. Within this frame, we have described the possibility of diffraction-free, superluminal, luminal and subluminal propagation in pulsed beams with a transverse structure explicitly governed by Fresnel diffraction in time. In comparison with \cite{KONDAKCI,ALONSO,KAMINER}, the new degree of freedom introduced by the pulse velocity makes it possible to have TD beams of arbitrary (paraxial) width for a given pulse duration. The simplicity of our approach eases the description of many other forms of TD beams, as Hermite-Gauss, vortex-carrying TD beams, etc. from well-established knowledge about monochromatic light beams, and establishes a basis for the study of their nonlinear propagation at the high intensities involved in their intended applications. We also offer a new perspective of the family of LWs \cite{SAARI}, of which TD beams is a broad subset, and of the luminal TD pulses in \cite{KAMINER}, stimulating ongoing research in modeling LWs for applications \cite{ZAMBONI}), the design of more sophisticated waves, as the nonparaxial tilted-phase-front beams recently introduced in \cite{WONG}, or their transformation by (paraxial) optical systems.

\end{document}